\documentclass[preprint,12pt,numbers,square,numbers]{elsarticle}
\usepackage{amssymb}
\usepackage{array}
\newcolumntype{P}[1]{>{\centering\arraybackslash}p{#1}}
\newcolumntype{M}[1]{>{\centering\arraybackslash}m{#1}}
\usepackage{a4wide,amssymb,epsfig,latexsym,multicol,array,hhline,fancyhdr,pictex, latexsym, graphicx,amsbsy,amsfonts,amsthm,verbatim}
\usepackage{supertabular}
\usepackage{amsmath}
\usepackage{lastpage}
\usepackage[lined,boxed,commentsnumbered]{algorithm2e}
\usepackage{enumerate}
\usepackage{color}
\usepackage{longtable}
\usepackage{arydshln}
\usepackage{graphicx}
\usepackage{array}
\usepackage{booktabs}
\usepackage{indentfirst}
\usepackage{tabularx, caption}
\usepackage{multirow}
\usepackage{multicol}
\usepackage{rotating}
\usepackage{graphics}
\usepackage{setspace}
\usepackage{epsfig}
\usepackage{tikz}
\usepackage{pgfplots}
\usepackage{array}
\usepackage{float}
\usepackage{wrapfig}
\usepackage{indentfirst}
\usepackage{comment}
\usepackage{tabularx}
\usepackage[most]{tcolorbox}
\usepackage{newunicodechar}
\usepackage{url}
\usepackage{hyperref} 

\usepackage{tikz-timing}[2014/10/29]
\usetikztiminglibrary[rising arrows]{clockarrows}
\usepackage{xparse}
\usetikzlibrary{arrows,backgrounds}
\usepackage{hyperref}
\usetikzlibrary{arrows,shapes.gates.logic.US,shapes.gates.logic.IEC,calc}
\hypersetup{urlcolor=blue,linkcolor=black,citecolor=black,colorlinks=true} 
\usepackage{footnote}
\usetikzlibrary{patterns}
\usepackage[]{algorithm2e}
\usepackage{xcolor}
\usepackage{listings}
\usepackage{color}
\usepackage{soul}
\usepackage{layout}
\usepackage{subcaption} 
\usepackage[pass,a4paper,margin=1in,bindingoffset=0.5in]{geometry}
\pgfplotsset{compat=1.18}
\newcommand{\PreserveBackslash}[1]{\let\temp=\\#1\let\\=\temp}
\newcolumntype{C}[1]{>{\PreserveBackslash\centering}p{#1}}
\newcolumntype{R}[1]{>{\PreserveBackslash\raggedleft}p{#1}}
\newcolumntype{L}[1]{>{\PreserveBackslash\raggedright}p{#1}}
\usepackage{listings}
\usepackage{color}

\definecolor{dkgreen}{rgb}{0,0.6,0}
\definecolor{gray}{rgb}{0.5,0.5,0.5}
\definecolor{mauve}{rgb}{0.58,0,0.82}

\definecolor{dkgreen}{rgb}{0,0.6,0}
\definecolor{gray}{rgb}{0.5,0.5,0.5}
\definecolor{mauve}{rgb}{01,0,0.82}
\lstdefinelanguage{Python}{
    keywords={def, return},
    keywordstyle=\color{blue},
    commentstyle=\color{gray},
    stringstyle=\color{green},
    identifierstyle=\color{black},
    basicstyle=\ttfamily\small,
    breaklines=true,
    showstringspaces=false,
    numbers=left,
    numberstyle=\tiny,
    numbersep=5pt,
    frame=tb,
    frameround=tttt,
    tabsize=4,
    columns=flexible,
    keepspaces=true
}

\usepackage{xcolor,pifont}
\newcommand*\colourcheck[1]{%
  \expandafter\newcommand\csname #1check\endcsname{\textcolor{#1}{\ding{51}}}%
}
\colourcheck{blue}
\colourcheck{green}
\colourcheck{red}

\NewDocumentCommand{\busref}{som}{\texttt{%
#3%
\IfValueTF{#2}{[#2]}{}%
\IfBooleanTF{#1}{\#}{}%
}}

\definecolor{darkgray}{rgb}{.4,.4,.4}
\definecolor{purple}{rgb}{0.65, 0.12, 0.82}

\lstdefinelanguage{js}{
  keywords={div, Button, img, Logo, Link, style},
  keywordstyle=\color{black}\bfseries,
  ndkeywords={class, export, boolean, throw, implements, import, const},
  ndkeywordstyle=\color{black}\bfseries,
  identifierstyle=\color{black},
  sensitive=false,
  comment=[l]{//},
  morecomment=[s]{/*}{*/},
  stringstyle=\color{black}\ttfamily,
  morestring=[b]',
  morestring=[b]"
}

\usetikzlibrary{arrows,backgrounds}
\usepackage{hyperref}
\hypersetup{urlcolor=blue,linkcolor=black,citecolor=black,colorlinks=true}

\begin{document}

\title{Evaluating Classical Software Process Models as Coordination Mechanisms for LLM-Based Software Generation}
\author[inst1,inst2]{Duc Minh Ha}
\author[inst3]{Phu Trac Kien}
\author[inst1,inst2]{Tho Quan}
\author[inst4]{Anh Nguyen-Duc}

\affiliation[inst1]{organization={Faculty of Computer Science and Engineering, Ho Chi Minh City University of Technology (HCMUT),  Vietnam}}
\affiliation[inst2]{organization={Vietnam National University Ho Chi Minh City,  Vietnam}}
\affiliation[inst3]{organization={FPT University, Vietnam}}
\affiliation[inst4]{organization={Department of Economics and IT, University of South Eastern Norway, Norway}}

\begin{abstract}
[Background] Large Language Model (LLM)-based multi-agent systems (MAS) are transforming software development by enabling autonomous collaboration. Classical software processes such as Waterfall, V-Model, and Agile offer structured coordination patterns that can be repurposed to guide these agent interactions. [Aims] This study explores how traditional software development processes can be adapted as coordination scaffolds for LLM-based MAS and examines their impact on code quality, cost, and productivity. [Method] We executed 11 diverse software projects under three process models and four GPT variants, totaling 132 runs. Each output was evaluated using standardized metrics for size (files, LOC), cost (execution time, token usage), and quality (code smells, AI- and human-detected bugs). [Results] Both process model and LLM choice significantly affected system performance. Waterfall was most efficient, V-Model produced the most verbose code, and Agile achieved the highest code quality, albeit at higher computational cost. [Conclusions] Classical software processes can be effectively instantiated in LLM-based MAS, but each entails trade-offs across quality, cost, and adaptability. Process selection should reflect project goals, whether prioritizing efficiency, robustness, or structured validation.
\end{abstract}

\maketitle

\section{Introduction}
The rise of agentic software engineering signals a potential paradigm shift in the way software systems are conceptualized, developed, and maintained \cite{rasheed_autonomous_2023,liu_large_2024}. AI agents powered by modern Large Language Models (LLMs) are no longer confined to passive roles; instead, they are becoming active participants throughout the software development lifecycle \cite{hong_metagpt_2024,rasheed_autonomous_2023}. Recent frameworks including AutoGen \cite{wu_autogen_2024}, MAS-GPT \cite{ye_mas-gpt_2025}, and MetaGPT \cite{hong_metagpt_2024} have demonstrated the feasibility of multi-agent configurations, where LLM-based agents can coordinate, negotiate, and iteratively construct a whole software product with various artifacts . 

When scaling the number of agents, the challenge of coordination—long central to software engineering—takes on renewed significance. In human-centric software development, coordination enables individuals with specialized roles to reduce complexity, manage interdependencies, and traceable development process \cite{humphrey_managing_1989,cataldo_socio-technical_2008,meidan_measuring_2018}. Traditional processes, such as Waterfall, V-Model and Agile embody this logic in different sequences and hierarchies \cite{nguyen_duc_dispersion_2012, nguyen_duc_forking_2014}. A similar imperative arises in multi-agent systems (MAS), where autonomous agents—each with partial knowledge and distinct capabilities—must rely on robust coordination mechanisms to align their efforts and achieve shared objectives. Current research on MAS highlights several practical limitations that hinder effective coordination, including agents’ restricted working memory, partial observability of the shared environment, and the high cost or limited efficiency of cross-agent communication \cite{sun_multi-agent_2025, agashe_llm-coordination_2025}. 


We argue that classical software engineering processes—such as Waterfall and Agile—can meaningfully enhance coordination in large-scale multi-agent systems (MAS), particularly those built on LLMs. These human-centric processes offer well-defined roles, structured workflows, and intermediate artifacts that promote transparency, interpretability, and control—qualities essential for managing agent behavior at scale. By empirically studying how MAS adopt and adapt elements like user story breakdowns, sprint planning, design documentation, and retrospectives, we can gain insight into how agentic collaboration unfolds and where human oversight is most beneficial. Artifacts such as design documents, test plans, and issue trackers serve not only as coordination checkpoints but also as promptable, memory-friendly communication layers between agents. Moreover, the modular nature of these processes enables varying levels of automation and human-in-the-loop involvement, supporting hybrid systems that are both scalable and auditable. An empirical study in this space would provide foundational knowledge for designing agent coordination mechanisms that are not only effective, but also interpretable, resilient, and aligned with established engineering practice.

This study investigates how software development processes can be used to coordinate LLM-based MAS in end-to-end development projects and how such alignment influences project performance. Our Research Questions (RQs) are:
\begin{itemize}
    \item RQ1. How can software processes (Waterfall, V-Model, Agile) be instantiated as coordination mechanisms for LLM-based multi-agent systems?
    \item RQ2. What differences arise in size, cost, and quality when these process-driven MAS configurations are applied to software projects?
\end{itemize}

We address these RQs by conducting experiments using the MetaGPT framework—a role-based, multi-agent LLM platform—across 11 software development projects. Each project is executed using one of the three process models, enabling systematic comparisons across coordination strategies. The contribution of this work is two-fold:
\begin{itemize}
\item We bridge traditional software engineering knowledge and modern multi-agent system paradigms by implementing human-centric processes—such as Agile and Waterfall—in LLM-based agent systems.
\item We enhance understanding of LLM-based multi-agent systems through an in-depth, multi-faceted empirical evaluation across various system configurations and practical software project contexts.
\end{itemize}
 
\section{Related Works}
This section presents background on coordination in software development proccesses (Section 2.1), current research on coordination of LLM-based MAS (Section 2.2) and the framework MetaGPT (Section 2.3).
\subsection{Coordination in different software processes}

Among the most established software process models are Waterfall, V-Model, and Agile methodologies \cite{sommerville_software_2016}. In the Waterfall model, coordination primarily occurs within distinct project phases. For example, a requirements analyst communicates with the customer during the requirements phase. This is followed by interactions between the requirements analyst and software designer during the design phase. In the development phase, software developers coordinate with designers and testers, and in the testing phase, testers engage with developers to resolve issues. Each phase depends on the successful transfer of information and outcomes from the previous phase, making sequential communication critical \cite{sommerville_software_2016}.

In the V-Model, coordination is tightly linked to verification and validation activities. For each development phase, there is a corresponding testing phase that requires alignment. For instance, while defining requirements, the team must simultaneously coordinate with testers to outline system testing plans. Similarly, during the design phase, the software designer collaborates with developers and testers to create integration testing plans. Throughout the project, feedback loops between development and testing phases ensure that outputs are verified and validated in parallel, emphasizing structured and cross-functional communication \cite{sommerville_software_2016}.

In the Agile model, coordination is iterative, continuous, and collaborative \cite{aasheim_process_2018,berntzen_taxonomy_2022}. Unlike the linear approaches of Waterfall and V-Model, Agile fosters real-time communication among all stakeholders, including customers, developers, testers, and designers. Agile ceremonies such as daily stand-ups, sprint planning, and retrospectives facilitate constant collaboration. Developers, testers, and designers work together within the same iteration, allowing for quick feedback loops and adjustments to meet evolving requirements. Coordination in Agile is more dynamic and relies on shared ownership and teamwork \cite{dyba_empirical_2008}.

As these models present three distinct approaches to coordination—sequential (Waterfall), verification-driven (V-Model), and iterative (Agile), we select them to investigate how such differing coordination logics influence software implementation within multi-agent systems.
\subsection{Coordination of LLM-based agent systems }
\begin{figure}[h]
  \centering
  \includegraphics[width=0.6\textwidth]{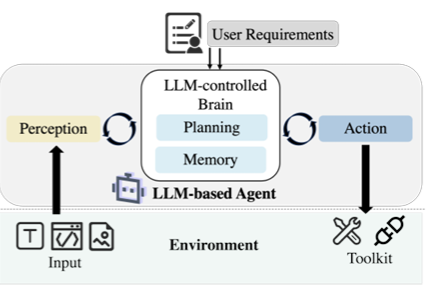}
  \caption{LLM-based agent system \cite{liu_large_2024}}
  \label{fig:LLMagentbasic}
\end{figure}

Large Language Model (LLM) Multi-Agent Systems are an emerging approach in the software engineering domain, where multiple LLM-based agents are assigned specific roles to collaboratively execute complex software development tasks. Instead of relying on a single general-purpose model, these systems divide responsibilities among specialized agents—such as product manager, system architect, programmer, and quality assurance tester—each designed to emulate a real-world role within a software engineering team. By coordinating through structured workflows and shared memory, the agents can interpret requirements, design architectures, write code, generate tests, and iterate on feedback. (see Figure \ref{fig:LLMagentbasic}). He et al. \cite{he_llm-based_2024} present a systematic review of LLM-based MAS in software engineering. Their work highlights the core affordances of such systems—autonomous task execution, mutual verification among agents, and improved scalability through agent specialization. They identify persistent challenges, such as trustworthiness and coordination complexity, 

A growing body of research has begun to explore multi-agent orchestration for end-to-end software generation. Guo et al. \cite{DBLP:Wu-et-al/2023a} summarize common interaction patterns among LLM-based agent systems, namely cooperative, debate, and competitive paradigms. Cooperative settings involve agents working collectively towards a shared solution, while debate and competitive structures encourage agents to critique or even oppose one another to improve solution quality. Beyond interaction modes, He et al. \cite{he_llm-based_2024} synthesized four typical communication architectures—layered, decentralized, centralized, and shared message pools—each with distinct trade-offs in efficiency and scalability. For example, MetaGPT operationalizes a shared message pool, a mechanism that maintains a shared message pool where agents publish messages and subscribe to relevant messages based on their profiles, hence,  reduces communication overhead and improves synchronization across roles \cite{hong_metagpt_2024}.  

Communication in these systems is generally realized through natural language interfaces, which support not only the exchange of task-relevant information but also higher-order interactions such as deliberation, clarification, and conflict resolution \cite{DBLP:Li-et-al/2023}. This has led to proposals for agent-oriented prompting languages to reduce ambiguity and improve reliability. Nevertheless, natural language orchestration introduces risks of inconsistency and “hallucination,” underscoring the need for stronger coordination frameworks.

More relevant to our work, some recent studies have begun exploring the feasibility of applying process logic to agent orchestration. ChatDev \cite{qian_chatdev_2024} demonstrates how conversational agents can collaboratively develop software through structured dialogues, but it primarily showcases feasibility rather than systematically comparing processes. AgileCoder \cite{nguyen_agilecoder_2024} introduces an Agile-inspired orchestration of LLM agents, highlighting the benefits of iteration and sprint-like cycles, but it does not benchmark Agile against alternative processes. Similarly, SWE-Bench provides a task-based benchmark for evaluating LLMs on software engineering problems, yet it focuses on model-level performance rather than process-level coordination \cite{jimenez_swe-bench_2024}.  

Our contribution extends this literature in two ways. First, instead of proposing a single orchestration paradigm, we operationalize \emph{three classical software processes} (Waterfall, Agile, and V-Model) as coordination scaffolds (comparing to one approach in AgileCoder \cite{nguyen_agilecoder_2024}). This provides a unified lens to examine how established process traditions map to LLM-based MAS and their coordination needs. Second, we go beyond conceptual demonstration by conducting one of the first \emph{systematic empirical evaluations} of process-driven coordination in LLM agent teams, analyzing cost, productivity, and quality trade-offs.  


\subsection{MetaGPT - a framework for collaborative AI agents}

Multi-Agent Systems (MAS) are computational systems composed of multiple interacting agents, each capable of autonomous decision-making and goal-directed behavior. These agents can cooperate, coordinate, or compete to solve complex problems that are difficult for a single agent or monolithic system to handle alone. MetaGPT, a well-known representative of MAS systems, operationalizes software processes by simulating a virtual software team in which LLM-based agents collaborate according to predefined roles and workflows \cite{DBLP:Hong-et-al/2023}. Each agent is assigned a specialized role—such as requirements analyst, architect, programmer, or tester—mirroring the division of labor in classical software engineering. Coordination is achieved through the exchange of structured artifacts, including requirement specifications, architecture diagrams, test reports, and source code. This artifact-driven workflow parallels documentation-heavy processes like Waterfall and the V-Model, where intermediate deliverables ensure traceability and alignment across stages. At the same time, the use of Standard Operating Procedures (SOPs) to guide task execution provides an Agile-like rhythm, in which agents progress iteratively through analysis, design, implementation, testing, and deployment. Thus, MetaGPT does not adhere strictly to one process model but blends elements from multiple traditions—sequential handoffs from plan-driven methods with iterative task cycles reminiscent of Agile practices.

Communication is orchestrated through a shared message pool designed as a publish–subscribe mechanism. This allows agents to 'publish' outputs and 'subscribe' to relevant inputs without the overhead of constant peer-to-peer communication. By decoupling agents in this way, MetaGPT achieves both scalability and synchronization: agents can operate independently on their assigned tasks yet remain coordinated within the overall workflow. This design reflects a key principle of software processes—balancing autonomy and interdependence—while making coordination interpretable, auditable, and efficient. In effect, MetaGPT demonstrates how classical software process structures can be reinterpreted as coordination frameworks for autonomous AI teams.

\section{Research Method}
\subsection{Study design}
This research adopts an exploratory case study approach to investigate how coordination processes can be applied in multi-agent systems within a real-world software development contex \cite{yin_2016}. The selected case, BVN, is a leading software and IT services provider with over 1,500 employees, operating as a wholly owned subsidiary of Robert Bosch GmbH. Founded in 2010 in Ho Chi Minh City, BVN delivers engineering solutions across embedded systems, automotive software, enterprise IT, and digital transformation. The company supports international clients through end-to-end project execution, employing both traditional (Waterfall, V-Model) and Agile development processes based on project requirements. Known for its process maturity and engineering quality, BVN plays a key role in Bosch’s global software ecosystem.

The overall workflow is illustrated in Figure 1. We begin by creating a configuration that combines specific software process models with selected LLM models. Each agent role (e.g., Project Manager, Developer, Tester) is implemented using tailored prompt templates that align with its responsibilities. The coordination logic is executed with predefined Standard Operating Procedures (SOPs) and customized according to the structure of each process model—Waterfall, V-Model, and Agile. Project outcomes are then evaluated and compared across configurations in terms of output size, development cost (measured by token usage), and quality.

The full implementation code, including agent configurations, process logic, and evaluation scripts, is available in our public GitHub repository\footnote{\url{https://github.com/anhn/swpro_agents}} for full reproducibility. Additionally, the complete analyzed dataset used in this study is also publicly accessible\footnote{\url{https://doi.org/10.5281/zenodo.17137668}}.

\subsection{Experiments setting}

We adopt a tailored quasi-experimental design approach to examine the effect of software process models (Waterfall, V-Model, Agile) on the performance of LLM-based MAS. In line with best practices from empirical software engineering and social intervention research \cite{shadish_experimental_2002,kampenes_systematic_2009,basili_empirical_2007}, we carefully constructed the experiment using four key design elements: non-randomized assignment, within-subject replications, pretest equivalence, and multiple experimental groups. Our research subject is software projects. Our experimental unit is a configuration (Project x Process Model x GPT model), each is executed in a disctinct run. This ensures each project scenario acts as its own control, thus neutralizing project-level variability. Furthermore, a pre-experimental baseline was established using uniform input prompts and SOPs across all trials, and we retained consistent agent roles and project specifications throughout to control for confounding task factors. Although agents are not human subjects, the project scenarios themselves serve as quasi-subjects in a repeated measures setup.

While random assignment of treatments is not feasible in this agent-based setting, we have minimized plausible alternative explanations for observed effects by embedding replication, control variables, and structured variation. We employed several experimental groups (i.e., different GPT models per process condition), allowing us to assess whether observed differences are attributable to the coordination model rather than model architecture alone. To further validate causal inference, we included nonequivalent dependent variables, such as output size (lines of code, number of files), which are expected to correlate with process complexity but not necessarily with output quality. The absence of correlation between these variables and quality outcomes strengthens the claim that quality differences arise from coordination logic rather than project scale.

\begin{figure}
    \centering
\includegraphics[width=0.7\linewidth]{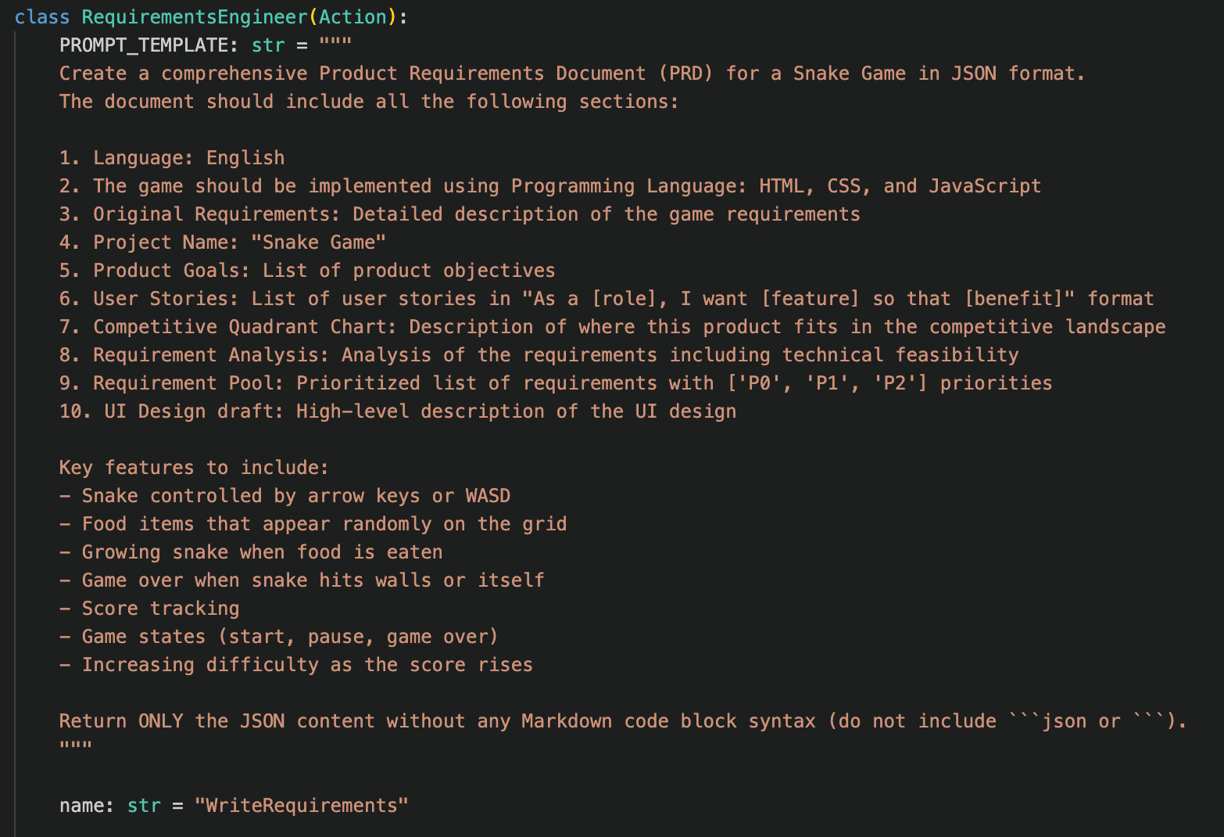}
    \caption{A sample requirement for a project}
    \label{fig:enter-label}
\end{figure}

Table \ref{tab:projects_table} describes the 11 software projects in this study. Each project follows a structured workflow that includes producing a requirement document, defining technical specifications, designing system architecture, implementing and testing code, and deploying final outputs, while evaluating critical metrics such as code quality, functional correctness, autonomy, and cross-project learning.
Each project starts with a piece of requirement description, as shown in Figure \ref{fig:enter-label}. The requirements were made to simulate the level of details given in early-stage real-life projects at BVN.

Each software project in the study was executed twelve times, representing all combinations of three software process models (Agile, Waterfall, V-model) and four GPT-based language models: GPT-4o-mini, GPT-4.1-nano, DeepSeek-Chat, and DeepSeek-Reasoning. To isolate the effect of process model structure from language model variability, each LLM was consistently assigned to the same set of agent roles across all process conditions. The prompt templates used for these roles are included in the Appendix. All experimental runs were performed using a standardized set of baseline requirements and identical agent configurations to ensure comparability. Table \ref{tab:quasi_design_summary} presents the nutshell of our experiment. The quasi-experiment was conducted between January and May 2025.
\begin{table}[ht]
\centering
\caption{Summary of Quasi-Experimental Design}
\begin{tabular}{|p{4.5cm}|p{10cm}|}
\hline
\textbf{Design Element} & \textbf{Description} \\
\hline
Design type & Quasi-experiment, within-subject replications across \\
\hline
Experimental unit & Project × Process Model × GPT Model \\
\hline
Intervention & Process model: Agile, Waterfall, and V-model. \\
\hline
Dependent variables & Size (LoC and files), Cost (resource usage and execution time), Quality (manual test) \\
\hline
Control measures & Uniform prompt templates and SOPs across all experiments; consistent agent role definitions; same LLM settings. \\
\hline
Statistical test & One-way ANOVA \\
\hline
\end{tabular}
\label{tab:quasi_design_summary}
\end{table}

\begin{figure}
    \centering
\includegraphics[width=0.6\linewidth]{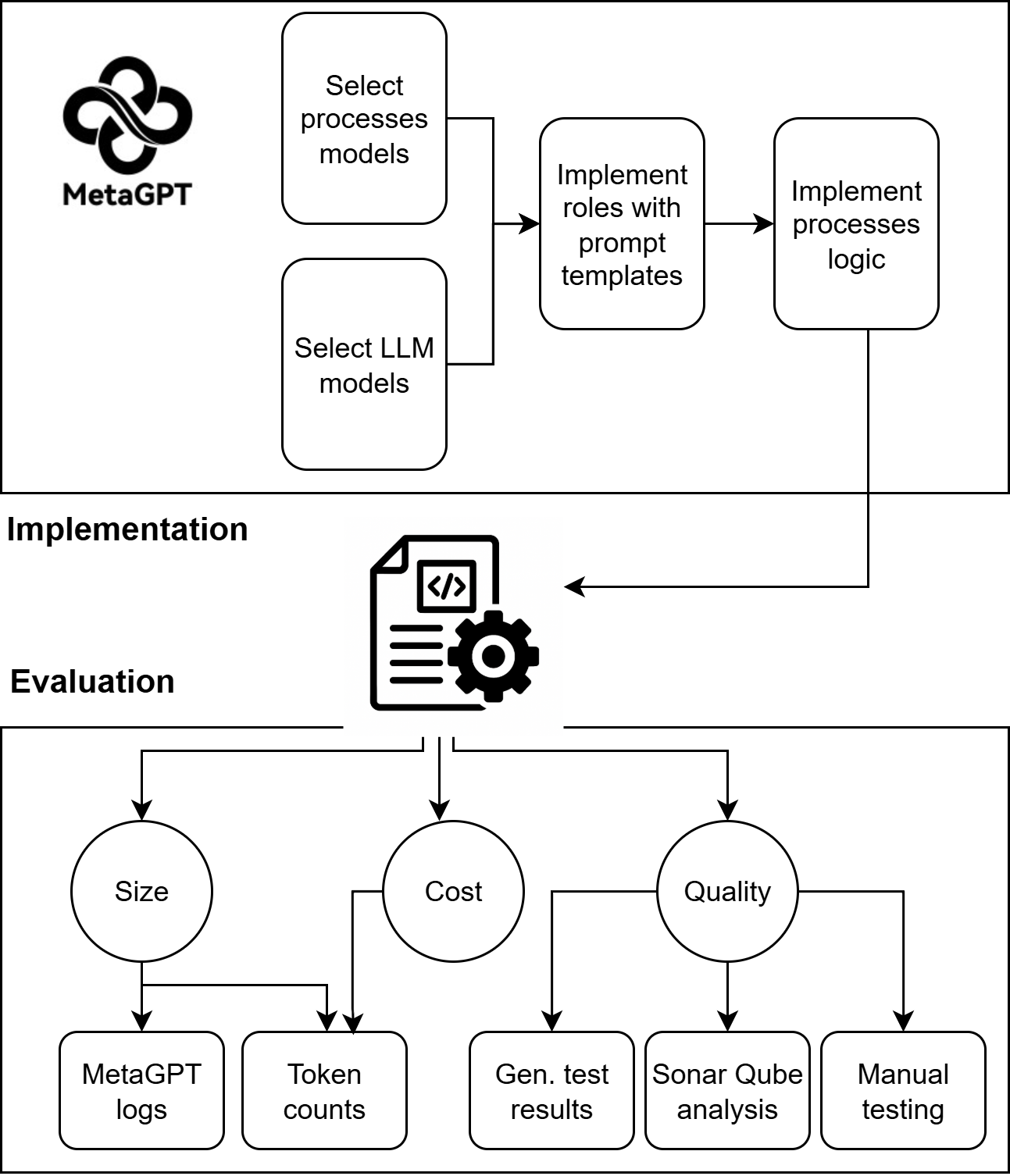}
    \caption{Overall workflow in this paper}
    \label{fig:exp-approach}
\end{figure}

\begin{table}[h]
\centering
\begin{tabular}{|p{4cm}|p{6cm}|p{2cm}|}
\hline
\textbf{Project} & \textbf{Description} & \textbf{Language} \\ \hline
Snake Game & Classic snake game with score tracking & JavaScript \\ \hline
Ping Pong Game & Two-player paddle game & JavaScript \\ \hline
Expense Tracker Application & Personal finance management tool & JavaScript \\ \hline
Hangman Game & Word guessing game & JavaScript \\ \hline
Space Invaders Game & Classic arcade shooter game & JavaScript \\ \hline
The Lord of The Rings Text-Based Game & Interactive fiction adventure & JavaScript \\ \hline
Flappy Bird Game & Side-scrolling obstacle avoidance game & Python \\ \hline
Excel Processor Application & Data manipulation for spreadsheets & Python \\ \hline
Tetris Game & Classic tile-matching puzzle game & Python \\ \hline
Learning English Application & Language learning tool & Python \\ \hline
QR Code Generator Application & Utility for creating QR codes & Python \\ \hline
\end{tabular}
\caption{Software Projects Overview}
\label{tab:projects_table}
\end{table}

\begin{table}[h]
  \centering
  \renewcommand{\arraystretch}{1.4} 
  \begin{tabular}{|l|p{11cm}|}
    \hline
    \textbf{Dimension} & \textbf{Operationalization} \\ \hline

    \textbf{Quality} &
    $Q_1 =$ Number of code smells (SonarQube) \\
    & $Q_2 =$ Number of vulnerabilities (SonarQube) \\
    & $Q_3 = \dfrac{\text{Bugs (AI detected)}}{\text{Number of test cases}}$ \vspace{5pt} \\
    & $Q_4 = \dfrac{\text{Bugs (Human detected)}}{\text{Number of test cases}}$ \vspace{5pt} \\ \hline

    \textbf{Cost} &
    $C_1 =$ Total tokens used \\
    & $C_2 =$ Total LLM execution time \\ \hline

    \textbf{Size} &
    $S_1 =$ Number of files \\
    & $S_2 =$ Total lines of code (LOC) \\
    & $S_3 = \dfrac{\text{Total tokens}}{\text{Total LOC}}$ \vspace{5pt} \\ \hline

  \end{tabular}
  \caption{Evaluation metrics}
  \label{tab:corrected_table}
\end{table}

\subsection{Evaluation metrics}
To assess software development outcomes in LLM-based multi-agent systems, we employ a comprehensive evaluation framework spanning quality, cost, and productivity dimensions (Table~\ref{tab:corrected_table}). The quality dimension includes metrics such as the number of code smells (indicating code maintainability), the number of vulnerabilities (capturing potential security issues), and the number of bugs, differentiated by whether they are detected by AI or human reviewers. The cost dimension accounts for multiple factors: the perceived complexity of AI agent configuration, the implementation time measured by LLM runtime, the computational cost of running the LLM, and the human effort needed for post-generation revision, approximated through time spent. Lastly, productivity is measured by both output size—captured by the number of files and total lines of code—and LLM effectiveness, calculated as the ratio of tokens used to lines of code produced.

\section{Results}

\subsection{RQ1 - How can software processes like Waterfall, V-Model, Agile be adapted and established within a context involving AI agents?}

In this section, we describe how MAS is applied to three popular software development methodologies, including Agile, Waterfall, and V-Model.

\subsubsection{Waterfall-like agent structure}
\begin{figure}
    \centering
    \includegraphics[width=0.4\linewidth]{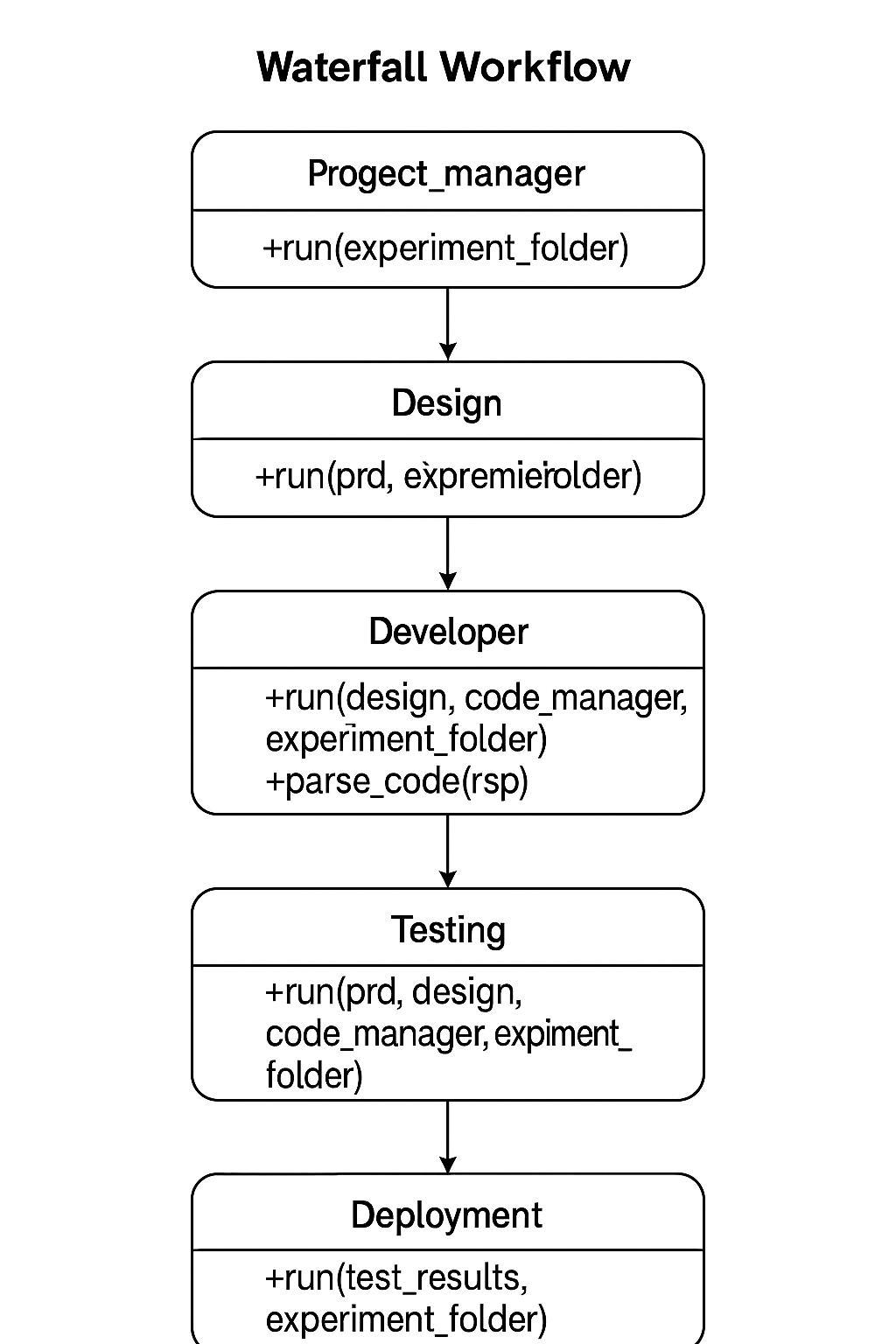}
    \caption{Waterfall Process}
    \label{fig:waterfall1}
\end{figure}

The Waterfall implementation in MetaGPT follows a strictly sequential pipeline with agents for Project Manager, Designer, Developer, Tester, and Deployer (Figure \ref{fig:waterfall1}. Coordination is layered and centralized, with the Developer acting as the main orchestrator, managing execution order and shared artifacts via CodeManager. Agents exchange information through structured natural language prompts encoding requirements, design, and outputs. Testing is separated into planning and execution roles (Unit, Integration, Acceptance), supporting specialization. However, the absence of backward feedback makes the system stable but less responsive to emergent errors.

\subsubsection{Agile-like agent structure}

\begin{figure}
    \centering
    \includegraphics[width=0.7\linewidth]{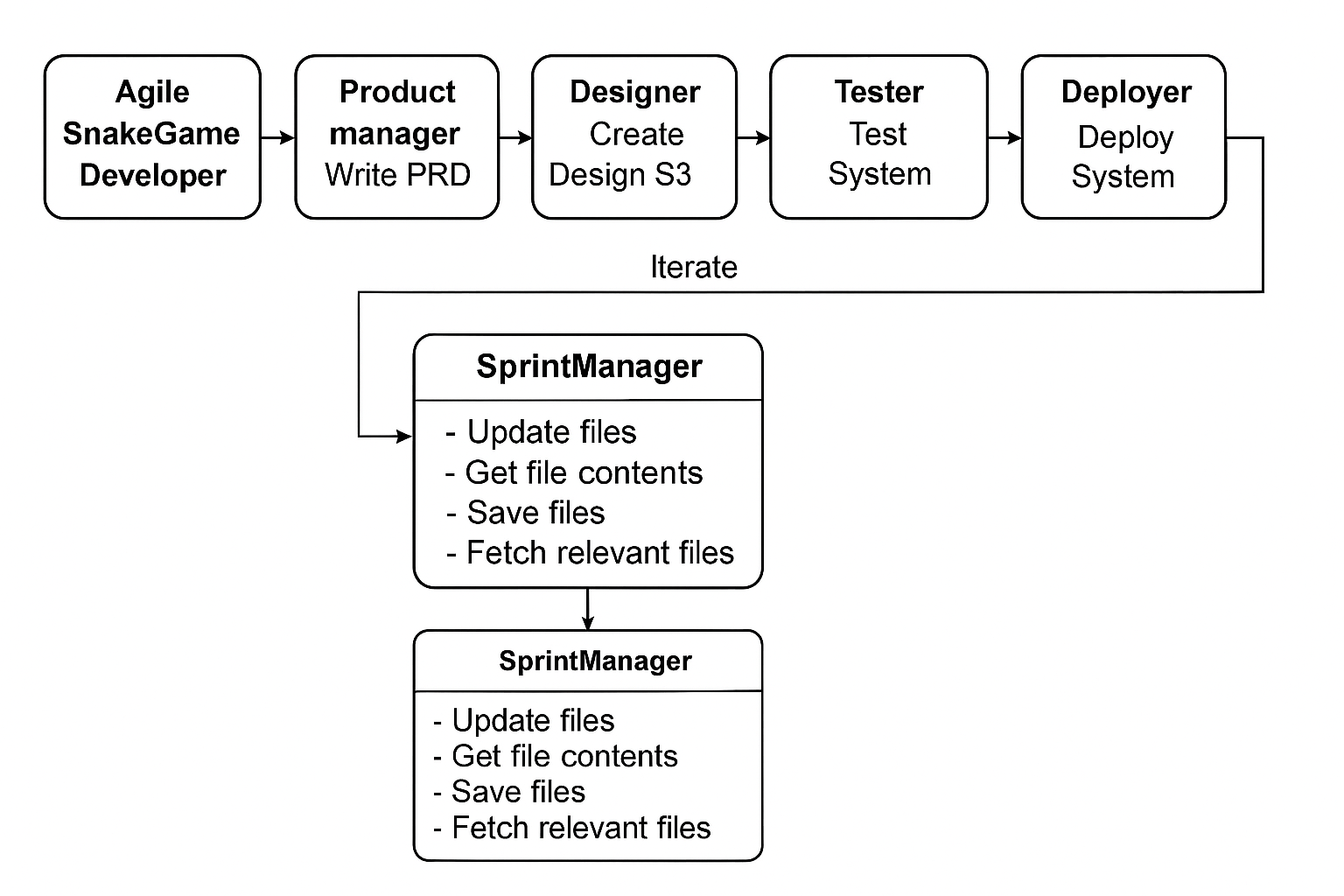}
    \caption{Agile Process}
    \label{fig:agile1}
\end{figure}

The Agile process (Figure \ref{fig:agile1}) emphasizes flexibility through iterative sprints, each covering requirements, design, implementation, testing, and deployment. Agents include Project Manager, Sprint Manager, Designer, Developer, Tester, and Deployer. The SprintManager coordinates sprint context, caching, and versioning, while the Developer centrally orchestrates agent execution. Shared artifacts such as requirements, design docs, and sprint code snapshots enable coordination. Iterative cycles allow greater adaptability and incremental progress, though central orchestration may introduce coordination overhead.

\subsubsection{V-model agent structure}
\begin{figure}
    \centering    \includegraphics[width=0.5\linewidth]{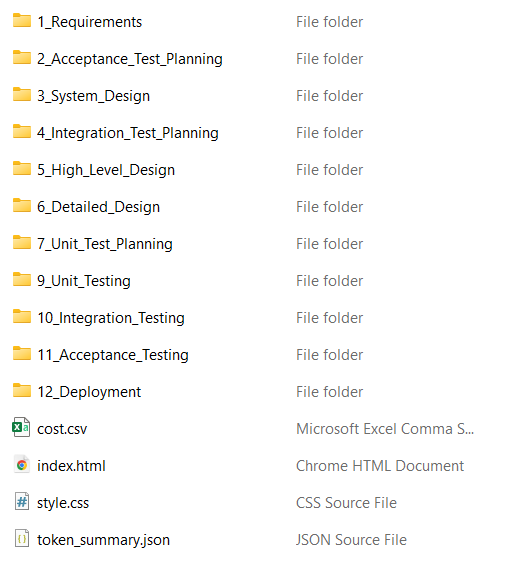}
    \caption{A Vmodel project structure}
    \label{fig:vprocess_folder}
\end{figure}

\begin{figure}
    \centering
    \includegraphics[width=0.4\linewidth]{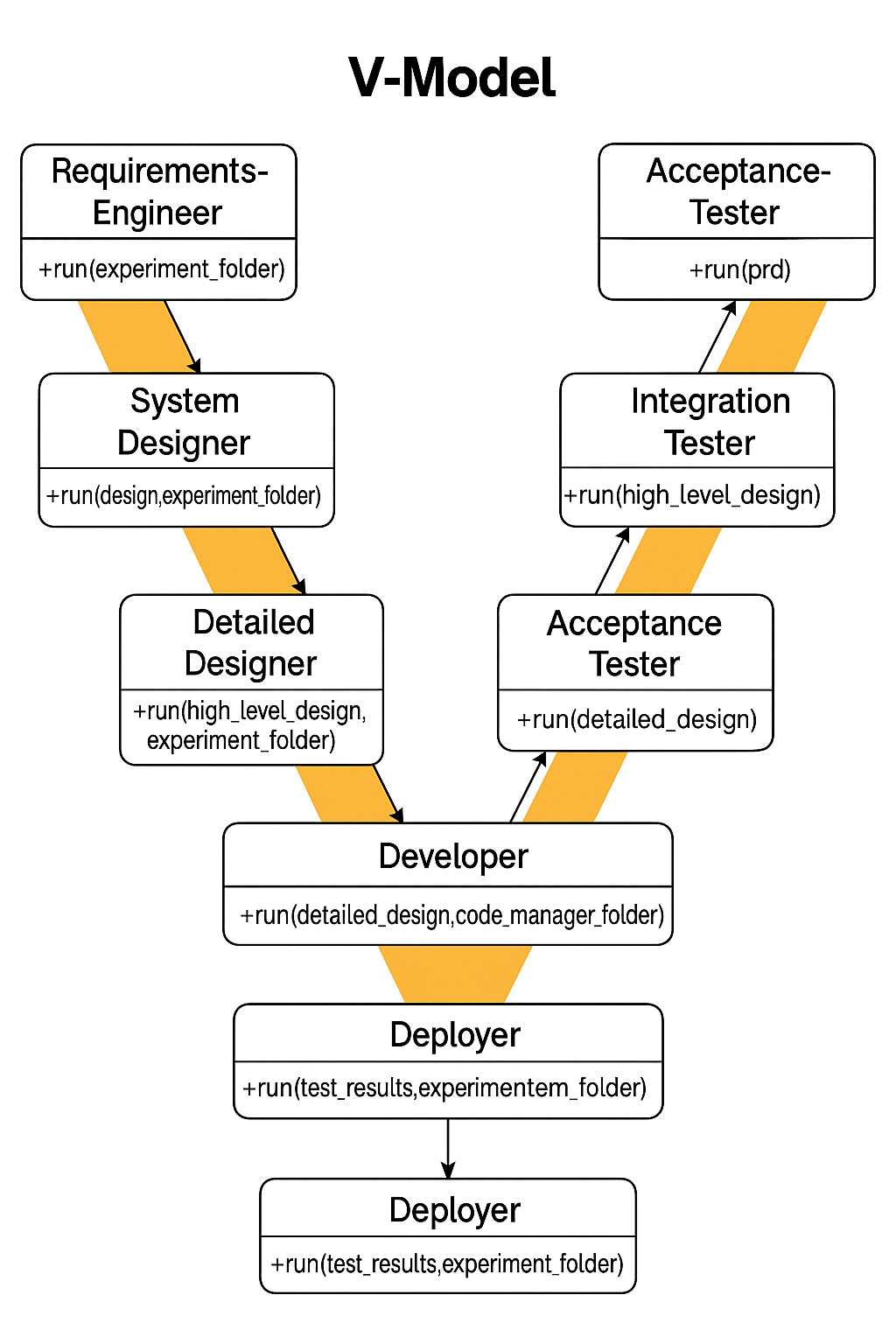}
    \caption{Vmodel Process}
    \label{fig:vprocess1}
\end{figure}

An example of a generated project folder is shown in FIgure \ref{fig:vprocess_folder}. The V-Model (Figure \ref{fig:vprocess1}) mirrors development and testing phases, pairing each stage (requirements, design, implementation) with a validation stage (acceptance, integration, unit testing). Agents include Project Manager, Designer, Developer, and three specialized test agents, with the Developer orchestrating phase transitions and managing artifacts via CodeManager. This structure enforces strict traceability and ensures validation maps directly to earlier specifications. While the approach strengthens alignment between tests and requirements, it limits adaptivity and provides minimal scope for negotiation or iteration. Appendix presents the list of prompts that have been used to configure a requirement analyst, a software designer, a tester and a developer in a Vmodel project. 
Table \ref{tab:process_comparison} summarizes the commonalities and differences among these models.

\begin{table}[H] 
  \centering
  \footnotesize
  \setlength{\tabcolsep}{6pt}
  \renewcommand{\arraystretch}{1.2}
  \begin{tabularx}{\textwidth}{|L{2.6cm}|X|X|X|}
    \hline
    \textbf{Aspect} & \textbf{Waterfall model} & \textbf{Agile model} & \textbf{V-model} \\ \hline

    \textbf{Agent roles} &
    Project Manager, Designer, Developer, Tester, Deployer (plus Unit/Integration/Acceptance testers separated) &
    Project Manager, Sprint Manager, Designer, Developer, Tester, Deployer &
    Project Manager, Designer, Developer, UnitTestExecutor, IntegrationTestExecutor, AcceptanceTestExecutor \\ \hline

    \textbf{Coordination structure} &
    Layered and centralized; Developer orchestrates sequential flow &
    Layered and centralized; SprintManager + Developer orchestrate sprint cycles &
    Layered and centralized; Developer enforces top-down and bottom-up sequence \\ \hline

    \textbf{Workflow style} &
    Strictly sequential; no backward feedback loops; phase handoffs only &
    Iterative sprints; small increments; sprint artifacts and snapshots maintained &
    Sequential with mirrored validation phases (requirements $\leftrightarrow$ acceptance testing, etc.) \\ \hline

    \textbf{Artifacts and communication} &
    Requirements, design, code, test reports; structured NL prompts; CodeManager handles artifacts &
    Requirements, design docs, sprint code snapshots; SprintManager handles caching, truncation, version control &
    Requirements, design, code; CodeManager manages file artifacts; validation tightly coupled to specifications \\ \hline

    \textbf{Testing approach} &
    Dedicated Tester role; later specialized executors (Unit, Integration, Acceptance) &
    Tester validates sprint outputs before deployment; sprint-based test coverage &
    Explicit test--development symmetry; each dev stage mapped to a specific testing stage \\ \hline

    \textbf{Adaptivity / feedback} &
    Limited to forward progression; stable but rigid &
    Iteration and feedback loops within sprints &
    Backward checks via test phases, but limited negotiation \\ \hline
  \end{tabularx}
  \caption{Comparison of MetaGPT implementations of Waterfall-like, Agile-like, and V-model-like processes}
  \label{tab:process_comparison}
\end{table}

\subsection{RQ2. What differences arise in size, cost, and quality when these process-driven MAS configurations are applied to software projects?}
\subsubsection{Comparing among Process models}
Table~\ref{tab:anova_results_stars} presents the results of one-way ANOVA tests assessing the impact of different software process models (Agile, Waterfall, V-model) on five key performance metrics. The analysis reveals statistically significant differences across process models for all metrics examined. In particular, the number of files and lines of code (LOC) show highly significant variation ($p < 0.01$), indicating that coordination style substantially influences the size of the generated codebase. Execution time and token cost are also significantly affected by process choice, with Agile and V-model generally incurring higher resource usage. Although the number of failed test cases detected by human testers shows a weaker effect ($p < 0.05$), it still suggests that process structure can influence code quality outcomes. The detail analysis for each performance metrics comes below.
\begin{table}[ht]
\centering
\caption{ANOVA results comparing the effect of Process Model on key performance metrics}
\begin{tabular}{|l|c|c|}
\hline
\textbf{Metric} & \textbf{F-Statistic} & \textbf{p-value} \\
\hline
Number of Files & 239.72 & $0.000\,^{**}$ \\
Lines of Code (LOC) & 67.16 & $0.000\,^{**}$ \\
Execution Time & 7.44 & $0.000\,^{**}$ \\
Token Cost & 17.45 & $0.000\,^{**}$ \\
Failure rate of manual tests & 3.27 & $0.041\,^{*}$ \\
\hline
\end{tabular}
\label{tab:anova_results_stars}
\end{table}
\subsubsection{Size}
In terms of files generated, the three process models exhibit distinct characteristics that reflect differences in coordination strategy and implementation complexity.
Table~\ref{tab:process_metrics} compares the number of files and total lines of code (LOC) per project. Agile projects generate the most files, with a median of 37 files per project—indicating more modularized and possibly more iterative task decomposition. Vmodel projects produce a moderate number of files (median = 19), while Waterfall yields the fewest, with a median of 13 files.

When considering code volume, Agile and Waterfall projects are more similar, with median LOC of 578 and 475, respectively. In contrast, Vmodel produces significantly larger codebases—with a median LOC of 934 and a maximum of 1,374—suggesting that its structured parallel workflows and extensive test planning may result in more verbose or redundant outputs. 

\begin{table}[h]
    \centering
    \begin{tabular}{|l|ccc|ccc|}
        \hline
        \textbf{Process} 
        & \multicolumn{3}{c|}{\textbf{No. of Files}} 
        & \multicolumn{3}{c|}{\textbf{No. of LOC}} \\
        \cline{2-7}
        & \textbf{Min} & \textbf{Med.} & \textbf{Max} 
        & \textbf{Min} & \textbf{Med.} & \textbf{Max} \\
        \hline
        Agile     & 11 & 37 & 54 & 236 & 578 & 1225 \\
        Waterfall & 10 & 13 & 20 & 253 & 475 & 791 \\
        Vmodel    & 13 & 19 & 19 & 610 & 934 & 1374 \\
        \hline
    \end{tabular}
    \caption{Comparison of project sizes across process models (number of files and lines of code)}
    \label{tab:process_metrics}
\end{table}

Not suprisingly, the scatter plot reveals a linear relationship between numbers of generated LOCs and total tokens used. Figure 7 also shows the clusters based on adopted development procceses. Notably, V-Model projects cluster at higher token and LOC values compared to Agile and Waterfall.

\begin{figure}
    \centering
    \includegraphics[width=0.5\linewidth]{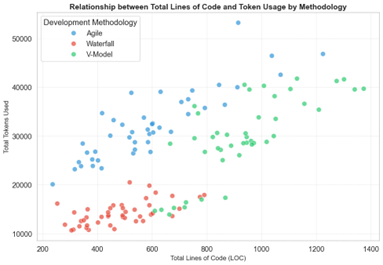}
    \caption{Total Used Tokens vs. Total Lines of Code}
\end{figure}

\subsubsection{Cost}
As shown in Table~\ref{tab:process_metrics}, the Waterfall model remains the most time-efficient among the three process models. It completes projects with a median execution time of 118 seconds, making it well-suited for scenarios with stable requirements and minimal iteration. The Vmodel, somewhat unexpectedly, demonstrates the second-fastest performance, with a median execution time of 226 seconds. This suggests that despite its emphasis on parallel validation phases, its static structure may allow agents to complete tasks without extensive looping or re-planning.

In contrast, Agile exhibits the highest execution time, with a median of 450 seconds and a maximum reaching 7,865 seconds. This wide range reflects Agile’s iterative nature, where agents frequently re-evaluate, regenerate, or refine outputs based on feedback. While more time-consuming, this behavior may contribute to improved robustness or alignment with evolving requirements, as discussed in earlier section

\begin{table}[h]
\centering
\caption{Comparison of execution time and total token usage across process models}
\begin{tabular}{|l|ccc|ccc|}
\hline
\textbf{Process} 
& \multicolumn{3}{c|}{\textbf{Execution Time (s)}} 
& \multicolumn{3}{c|}{\textbf{Total Tokens Used}} \\
\cline{2-7}
& \textbf{Min} & \textbf{Med.} & \textbf{Max} 
& \textbf{Min} & \textbf{Med.} & \textbf{Max} \\
\hline
Agile     & 71  & 450  & 7865  & 13,302 & 30,787 & 53,212 \\
Vmodel    & 74  & 226  & 2110  & 13,895 & 29,783 & 41,758 \\
Waterfall & 50  & 118  &  958  & 10,646 & 13,720 & 20,493 \\
\hline
\end{tabular}
\label{tab:process_cost}
\end{table}

Regarding token used, Waterfall demonstrates the most efficient token utilization, with a median of 16,142 tokens per project and ranging from 12,520 to 20,493 tokens. Agile methodology, by design, consumes a moderate volume of tokens, with a median of 30,408 tokens per project, ranging from 23,808 to 46,812 tokens. The V-Model emerges as the most token-intensive methodology, with a median of 39,630 tokens per project and ranging from 35,385 to 41,758 tokens. 

\begin{tcolorbox}[title=Observation 1, colback=gray!5, colframe=black, fonttitle=\bfseries]
The generated projects show clear differences in cost and size. Waterfall is the most efficient, with the lowest execution time, token usage, and file count. Vmodel produces the most lines of code, while Agile generates the most files and incurs the highest cost due to its iterative nature.
\end{tcolorbox}

\subsubsection{Quality}
\begin{figure*}
    \centering
    \includegraphics[width=1\textwidth]{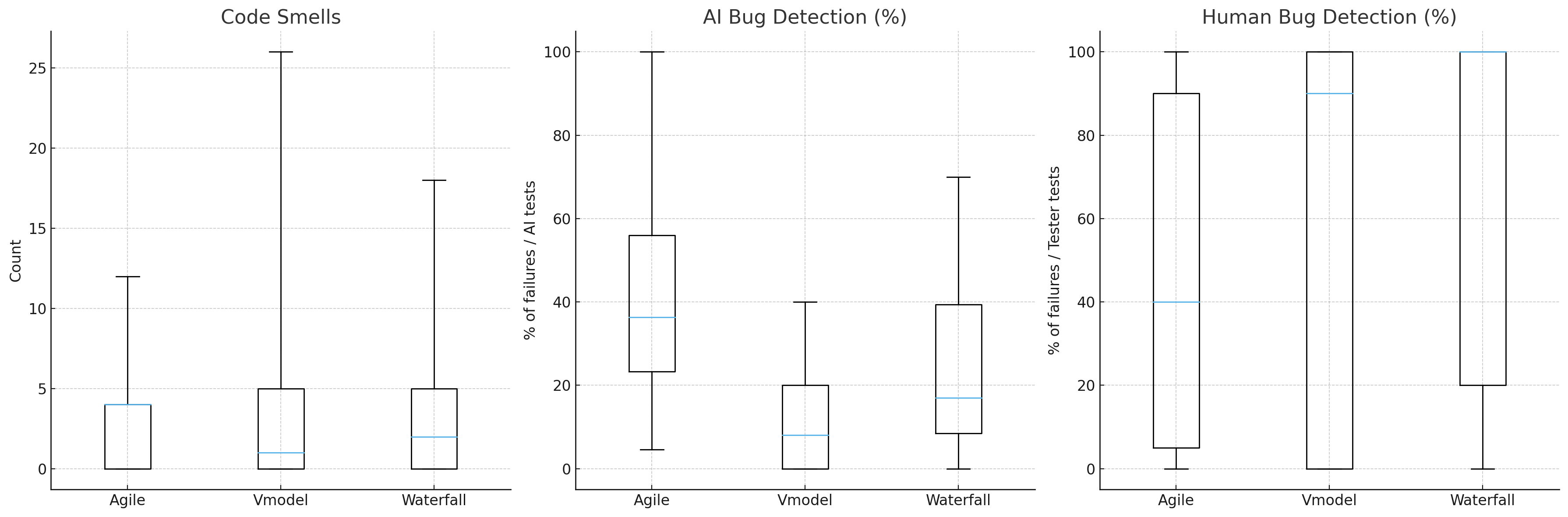}
    \caption{Code quality across software processes}
    \label{fig:comgptquality}
\end{figure*}

Figure~\ref{fig:comgptquality} compares code quality across Agile, Vmodel, and Waterfall configurations, using three indicators: the number of code smells, the percentage of bugs detected by AI-generated test cases, and the percentage of bugs identified through manual testing.

The first panel of the figure highlights the number of code smells found in generated projects. All three process models produced relatively clean code, with median values of 4 for Agile, 1 for Vmodel, and 2 for Waterfall. While Waterfall and Agile show compact interquartile ranges, Vmodel exhibits greater variability, with some projects showing extremely clean results (zero smells) and others with significantly higher counts. This inconsistency may reflect the Vmodel's structural rigidity, which—while intended to enforce discipline—can lead to mismatches between agent role execution and practical code generation, especially in the absence of adaptive feedback.

In the second panel, AI-driven bug detection rates are visualized. Agile shows the highest median bug detection rate at 35.6\%, followed by Waterfall at 18.5\%, and Vmodel at just 8.3\%. These findings are notable because they challenge assumptions about the Vmodel’s verification strength. Although Vmodel enforces a validation-first design, its static nature may impair the LLM agents' ability to generate executable and contextually relevant test cases. Without iterative refinement, test logic may remain superficial, resulting in fewer bug detections. In contrast, Agile’s continuous feedback and incremental structure likely enable agents to adapt their tests in response to observed behaviors—leading to greater alignment between bugs and test coverage.

The third panel presents the percentage of test failures found through human testing, offering a more grounded view of actual code reliability. Agile code exhibited the lowest failure rate, with a median of 40\%, while Vmodel and Waterfall showed significantly higher medians at 90\% and 100\%, respectively. This sharp contrast implies that while Vmodel and Waterfall are theoretically rigorous, their linear or mirrored workflows may not translate into practically reliable outputs when executed by autonomous LLM agents. The lack of iterative correction phases likely prevents early bug resolution. Agile’s lower failure rate suggests that its flexible coordination enables more responsive development, better aligning agent actions with functional expectations.

Together, these results underscore that in LLM-based agent systems, traditional process rigor does not guarantee higher code quality. Instead, adaptability, iteration, and feedback loops—core to Agile—appear to enhance both testability and functional correctness in autonomous agent workflows. Process models that rely on static plans and strict role separation may require additional scaffolding to perform effectively in generative, agentic environments.

\begin{tcolorbox}[title=Observation 2, colback=gray!5, colframe=black, fonttitle=\bfseries]
Agile consistently outperformed other models in quality metrics, with significantly higher AI bug detection and lower human-reported failure rates. This suggests that iterative coordination better supports LLM agents in producing robust and testable code.
\end{tcolorbox}

 \subsection{Comparing among LLM models}

Table~\ref{tab:anova_gpt_results} reports the results of ANOVA tests evaluating the effect of different GPT models on five performance metrics across software development tasks. The analysis shows that the choice of GPT model has a highly significant effect on lines of code (LOC), execution time, and token cost ($p < 0.01$), indicating that different LLM models lead to substantial differences in both code generation behavior and computational efficiency. In terms of execution efficiency, the GPT-based models clearly outperform the DeepSeek family. GPT-4.1 Nano executes in only 90 seconds on average, and GPT-4o Mini requires 145 seconds, whereas both DeepSeek Chat and DeepSeek Reasoner exceed 660 seconds. Token consumption follows a similar pattern: DeepSeek Chat produces the largest outputs with an average of 30,912 tokens, while DeepSeek Reasoner also remains high at 24,597 tokens. In contrast, GPT-4.1 Nano and GPT-4o Mini generate leaner responses at 27,047 and 19,041 tokens, respectively. These differences suggest that while the DeepSeek models tend to generate more verbose and resource-intensive outputs, the GPT models are more efficient in both runtime and token economy, making them preferable when responsiveness and throughput are prioritized.

\begin{table}[ht]
\centering
\caption{ANOVA results comparing the effect of GPT Model on software performance metrics}
\begin{tabular}{|l|c|c|}
\hline
\textbf{Metric} & \textbf{F-Statistic} & \textbf{p-value} \\
\hline
Number of Files & 0.047 & 0.9865 \\
Lines of Code (LOC) & 7.72 & $0.000^{**}$ \\
Execution Time & 15.79 & $0.000^{**}$ \\
Token Cost & 74.33 & $0.000^{**}$ \\
Failed Test Cases (by tester) & 2.38 & 0.0726 \\
\hline
\end{tabular}
\label{tab:anova_gpt_results}
\end{table}

\begin{figure}
    \centering
    \includegraphics[width=0.6\linewidth]{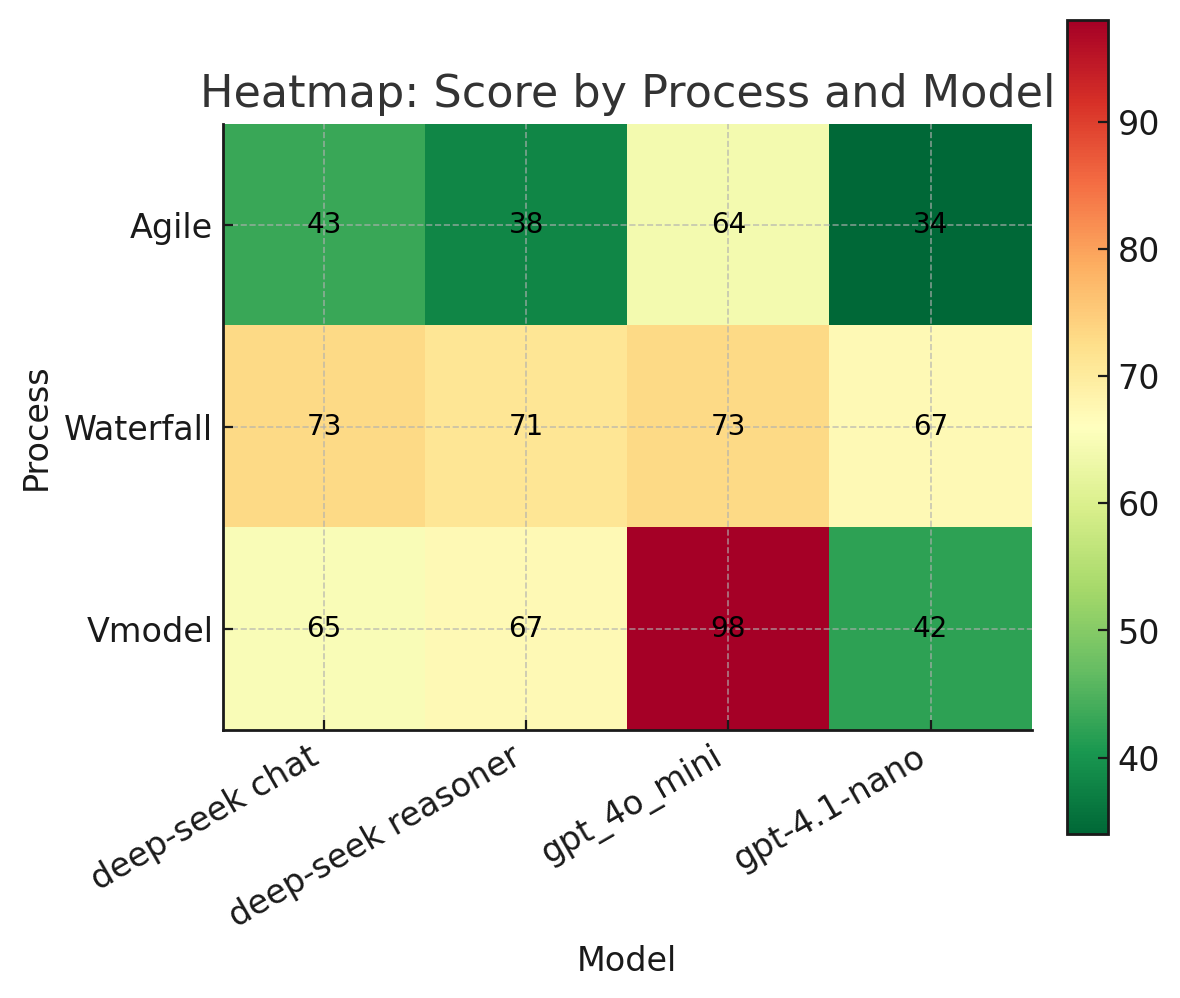}
    \caption{Code quality across software processes and LLMs}
    \label{fig:heatmap_comparison}
\end{figure}

In contrast, the number of files produced and the number of failed test cases identified by human testers are not significantly affected by the GPT model. Figure \ref{fig:heatmap_comparison} shows overall, the two deepseek variants achieve higher AI BDR compared to the GPT-based models. Deepspek reseasoner exhibits the strongest performance in AI bug detection, with an ratio of 54.62\%, considerably higher than Deepseek chat (30.39\%). In contrast, the GPT-based models (GPT 4o mini and GPT 4.1 nano) show markedly lower AI detection ratio, averaging 8.81\% and 22.17\%, respectively. The results in Figure \ref{fig:heatmap_comparison} implies that GPT-4.1-nano delivers the best overall code quality across the evaluated configurations. This might suggest that pairing lightweight models like GPT-4.1-nano with iterative coordination structures such as Agile may lead to more robust and reliable code output. In contrast, GPT-4o-mini performs the worst, particularly under the V-Model, where it reaches a score of 98, indicating significant quality degradation.

Interestingly, although GPT‑4.1-nano and GPT‑4o-mini are both positioned as smaller models, GPT‑4.1 (including nano/mini variants) supports up to 1 million input tokens, a substantial improvement over GPT‑4o’s 128K-token limit. This expanded context window is particularly beneficial in agentic software development, where models must reason over large codebases or long prompts involving multi-step plans. These findings align with previous benchmarks: GPT‑4.1 achieves 54.6\% on SWE‑bench Verified, representing an absolute improvement of 21.4\% over GPT‑4o \cite{nguyen_agilecoder_2024}.

\begin{tcolorbox}[title=Observation 3, colback=gray!5, colframe=black, fonttitle=\bfseries]
LLM model choice significantly impacted execution time, token usage, and code verbosity, but had limited influence on overall code quality.
\end{tcolorbox}

 \section{Discusison}
This section discusses the findings in Section 5.1, practical implications in Section 5.2 and threats to the validities in Section 5.3.
\subsection{Impact of Software Processes on LLM Agents}
The results support the conclusion that the choice of software process model significantly influences key performance outcomes in LLM-based multi-agent systems. On Observation 1, as shown in the ANOVA results, process model selection has a strong and statistically significant impact on core project characteristics such as code size (measured in number of files and LOC), execution time, and token cost. Among the three models, Waterfall consistently produced smaller outputs and required fewer tokens, suggesting greater efficiency. In contrast, V-Model led to significantly larger codebases, indicating higher implementation complexity and resource demands—likely due to its built-in emphasis on parallel development and validation phases

Regarding Observation 2, with regard to code quality, the effects were less pronounced. The number of failed test cases reported by human testers was only marginally affected by process model (p = 0.041), and not significantly influenced by GPT model selection. Therefore, while Agile configurations showed somewhat lower failure rates, the evidence does not strongly support the claim that Agile produced more functionally reliable code. Any such trend should be interpreted cautiously, as the observed differences in failure rates are not robustly supported by the statistical analysis.

The results support the idea that selecting different software process models meaningfully influences the behavior and outcomes of LLM-based agent frameworks. As shown in Observation 1, the choice of process model affects core project characteristics such as size and cost, with Waterfall yielding the smallest codebase and lowest token usage, and V-Model generating significantly larger outputs. This suggests that more structured and verification-heavy processes lead to greater implementation complexity and resource consumption.

In terms of quality, Observation 2 reveals that Agile-based coordination consistently results in more functionally reliable code. Agile configurations achieved both the highest AI-driven bug detection rates and the lowest failure rates during manual testing, indicating superior robustness in execution. This may be attributed to the iterative feedback loops inherent in Agile, which allow agents to revise and refine their outputs more effectively than in rigid pipelines.

Observation 3—that GPT model choice plays a critical role in quality—is not supported by the data. While the GPT model had a strong influence on token cost, execution time, and LOC, its effect on failed test cases was not statistically significant (p = 0.073). This suggests that LLM architecture affects efficiency and verbosity, but not necessarily correctness or quality, at least as measured through human-reported test failures.

Taken together, these findings highlight the importance of process model selection in shaping the performance of LLM-based agent teams, particularly in terms of cost and output size. Waterfall may be suitable for resource-constrained or well-scoped projects, while V-Model may support traceability and thoroughness—albeit at higher execution cost. Agile remains a promising coordination structure, especially in settings that require iterative refinement, but its superiority in quality outcomes remains tentative rather than conclusive. Ultimately, coordination strategies for LLM agents should be selected based on the specific trade-offs and constraints of the development context.

\subsection{Practical Implications}

The findings of this study offer several actionable insights for developers building LLM-based agent frameworks:
\begin{itemize}
    \item Choose Process Models Strategically: The coordination structure inherent in each process model has a measurable impact on output size, cost, and execution time. Teams should choose between Agile, Waterfall, or V-Model based on whether their project emphasizes adaptability, efficiency, or traceability
    \item Use Agile for Robustness and Quality: While Agile promotes iterative refinement, its impact on code quality was only marginally significant. However, its structure remains advantageous in contexts that require continuous validation or dynamic specification updates
    \item Beware of Overhead in Structured Models: Highly structured models like the V-Model may increase implementation overhead and token usage without clear gains in correctness. Developers should be cautious when applying rigid frameworks to autonomous LLM workflows, which may not benefit from traditional process granularity.
    \item Model Choice Affects Efficiency More Than Quality Differences across GPT models significantly affected LOC, execution time, and cost—but not failed test rates. Developers should prioritize model selection based on resource constraints and generation behavior, rather than assuming larger models yield better functional results.
    \item Optimize for Context and Resources: Waterfall proved the most resource-efficient, generating smaller codebases with lower token consumption. This makes it suitable for constrained environments. Agile and V-Model, while offering more iterative or parallel flows, incur greater computational cost.
    \item Design with Modularity and Flexibility: Defining standard operating procedures (SOPs) and structured agent roles enables more consistent coordination and easier troubleshooting. This modular approach supports scalability, debugging, and potential integration with human-in-the-loop workflows or domain-specific extension.  
\end{itemize}

\subsection{Threats to validities}
We recognize several potential threats to validity that may influence the interpretation and generalizability of our findings. In designing the study as a quasi-experiment, we adopted established techniques to strengthen internal, construct, and external validity, while acknowledging the inherent limitations of non-randomized, agent-based evaluations.

Internal Validity: To mitigate threats to internal validity—specifically selection bias and confounding variables—we used a within-subject replication design, applying all combinations of process model and GPT model to the same set of software project tasks. This allows each project to serve as its own control and isolates the effect of the treatment (i.e., process model) from project-specific characteristics. Agent role prompts, SOPs, and LLM configurations were held constant across runs to ensure consistency. However, the stochastic nature of LLM outputs introduces non-determinism. We addressed this by executing multiple independent runs per condition and using aggregated outcome metrics. While this approach reduces random variation, it does not eliminate all potential carryover or model-specific learning effects, especially in the absence of full randomization.

Construct Validity: concerns whether our measurement strategy accurately captures the theoretical concepts under study—namely, code quality, resource efficiency, and productivity. To improve construct validity, we operationalized each construct using multiple indicators. For example, code quality was measured through AI- and human-executed test failures, as well as static analysis metrics like code smells and reported vulnerabilities. This triangulation reduces reliance on any single proxy and follows recommended practices in software engineering research. Still, we caution that some metrics, such as lines of code (LOC) or number of files, are size-oriented proxies and should not be interpreted as direct indicators of productivity or maintainability \cite{fenton_software_2014}. Similarly, while SonarQube-derived code smells provide valuable signals, their relationship to long-term software health is contextual and not always straightforward \cite{moonen_code_2012}. We addressed this by combining quantitative metrics with human validation to better reflect meaningful quality concerns \cite{mendes_cross-_2014}.

External Validity: relates to how well the findings generalize beyond the study setting. Our experimental tasks involved eleven small-to-medium scale projects, including simple games and utility tools. These project types are aligned with those commonly used in LLM benchmarking studies (e.g., HumanEval \cite{chen_evaluating_2021}, MBPP \cite{austin_program_2021}), where the focus is on reasoning and correctness under controlled conditions. Such settings are methodologically justified for isolating the effects of coordination logic, but they do not reflect the full complexity of industrial-scale systems. In real-world environments, software development is shaped by evolving requirements, multi-team coordination, regulatory compliance, and long-lived codebases. Consequently, process advantages observed in this setting—for example, the relative speed of Waterfall or the adaptability of Agile—may manifest differently in enterprise-grade or safety-critical systems \cite{stol_abc_2018}.

To enhance realism and reduce tool bias, all experiments were conducted using the MetaGPT framework. This provided consistent agent orchestration and file management but may introduce platform-specific affordances that favor certain coordination models (e.g., Waterfall’s sequential handoffs may align better with MetaGPT’s built-in CodeManager). Cross-framework validation using alternatives like CAMEL or AutoGen remains a promising direction for future research \cite{hong_metagpt_2024}.

\section{Conclusions}
This paper investigated how classical software process models—Agile, Waterfall, and Vmodel—can be instantiated as coordination scaffolds for LLM-based multi-agent systems using the MetaGPT framework. In addressing RQ1, we showed that each process model can effectively structure agent workflows, but with distinct behavioral and performance characteristics: Waterfall enables compact, efficient execution; Vmodel emphasizes structural completeness; and Agile supports adaptive refinement through iterative coordination.

In answering RQ2, our empirical analysis revealed statistically significant differences in cost, size, and quality metrics across process models. Waterfall was the most efficient in both execution time and token usage, producing smaller, less complex outputs. Vmodel generated the most verbose code (highest LOC), while Agile incurred the highest computational cost but achieved superior code quality, with the highest AI bug detection and lowest human failure rates. In contrast, differences across LLM models were more pronounced in efficiency-related metrics (e.g., runtime and token use) than in code correctness, suggesting that process structure plays a more critical role in shaping quality outcomes than model architecture alone.

Future work will extend this line of inquiry by investigating alternative coordination paradigms such as debate, marketplace-based negotiation, or decentralized peer-to-peer interaction. We aim to examine how different coordination styles affect system interoperability, code consistency, and fault localization. Additionally, we plan to scale our experimental framework to larger, multi-module software systems with interdependent components, exploring how coordination mechanisms adapt to increased complexity, role specialization, and long-running development cycles. Such work will contribute to establishing generalizable orchestration principles for trustworthy, scalable, and collaborative AI software engineering ecosystems.

\section{Appendix}
\appendix
\section{Prompt Templates}

\subsection{Prompt Template for PRD}
\begin{lstlisting}
PROMPT_TEMPLATE: str = """
Create a comprehensive Product Requirements Document (PRD) for an Expense Tracker application in JSON format.
The document should include all the following sections:

1. Language: English
2. The application should be implemented using Programming Language: HTML, CSS, and JavaScript
3. Original Requirements: Detailed description of the application requirements
4. Project Name: "Expense Tracker"
5. Product Goals: List of product objectives
6. User Stories: List of user stories in "As a [role], I want [feature] so that [benefit]" format
7. Competitive Quadrant Chart: Description of where this product fits in the competitive landscape
8. Requirement Analysis: Analysis of the requirements including technical feasibility
9. Requirement Pool: Prioritized list of requirements with ['P0', 'P1', 'P2'] priorities
10. UI Design draft: High-level description of the UI design

Key features to include:
- Expense categories: Food, Transportation, Hotel
- Ability to add new expenses with amount, category, date and description
- Interactive UI with clickable elements
- Display of expense history
- Summary statistics by category
- Ability to edit and delete expenses
- Responsive design that works on mobile and desktop

Return ONLY the JSON content without any Markdown code block syntax (do not include ```json or ```).
"""
\end{lstlisting}

\subsection{Prompt Template for System Design}
\begin{lstlisting}
PROMPT_TEMPLATE: str = """ 
Design a complete system architecture for {project_name} based on the entire Product Requirements Document:
{prd}

Create a comprehensive design document in JSON format with these sections:
1. architecture_description: Text description of the overall architecture
2. class_diagram: Mermaid.js syntax for the class diagram
3. data_flow: Description of data flow between all components
4. ui_design: Detailed description of all UI components and their behavior
5. state_management: How all application states are handled
6. requirements_mapping: How each PRD requirement maps to this implementation

Following the V-Model methodology, your design must be complete and comprehensive, covering:
- Expense data model with amount, category, date and description
- Interactive UI components with click handlers
- Category management system
- Expense CRUD operations
- Data persistence using localStorage
- Summary statistics calculation
- Responsive layout implementation

Return ONLY the JSON content without any Markdown code block syntax (do not include ```json or ```).
"""
\end{lstlisting}

\subsection{Prompt Template for Testers}
\begin{lstlisting}
PROMPT_TEMPLATE: str = """
Create comprehensive acceptance test cases for the {project_name} based on the Product Requirements Document:
{prd}

Following the V-Model methodology, these acceptance tests should directly verify the requirements specified in the PRD.

Create a detailed acceptance test plan in JSON format with these sections:
1. test_cases: List of test cases with descriptions and expected results
2. requirements_coverage: Mapping of each requirement to test cases
3. pass_criteria: Criteria for determining if the system passes acceptance testing
4. test_environment: Description of the required test environment
5. test_data: Description of any required test data

Focus on testing:
- Expense creation with all required fields
- Category filtering and display
- Interactive UI elements
- Data persistence
- Summary statistics accuracy
- Responsive design

Return ONLY the JSON content without any Markdown code block syntax (do not include ```json or ```).
"""
\end{lstlisting}

\subsection{Prompt Template for Developers}
\begin{lstlisting}
PROMPT_TEMPLATE: str = """
Implement the complete {project_name} according to the detailed design:
{detailed_design}

Following the V-Model methodology, you must implement the entire system according to the specifications.

Create all these files:
- index.html
- style.css
- script.js

For an Expense Tracker, ensure you implement:
- Complete expense management with amount, category, date and description
- Three main categories: Food, Transportation, Hotel
- Interactive UI with clickable elements
- Data persistence using localStorage
- Expense history display
- Summary statistics by category
- Ability to edit and delete expenses
- Responsive design that works on mobile and desktop

The UI must be interactive with clear click handlers for all actions.

Return the complete code for all files in the following format:
```html
<!-- index.html -->
<your_complete_html_code_here>
"""
\end{lstlisting}

\bibliographystyle{unsrt}
\bibliography{library}

\end{document}